\documentclass[twoside]{article}

\usepackage{aistats2024}
\usepackage{natbib}
\usepackage{subfig}
\usepackage{amssymb}
\usepackage{graphicx}
\usepackage{mathtools}
\usepackage{url}

\newcommand{\mat}[1]{{\mathbf{{#1}}}} 
\newcommand{\bvec}[1]{{\mathbf{#1}}}
\newcommand{\trsign}{{\mathsf{T}}}
\newcommand{\tr}{^{\trsign}}

\newcommand{\bP}[0]{\mat{P}}

\newcommand{\bS}[0]{\mat{S}}

\newcommand{\R}[0]{\mat{R}}

\newcommand{\x}[0]{\bvec{x}}

\newcommand{\y}[0]{\bvec{y}}

\newtheorem{theorem}{Theorem}
\newtheorem{prop}[theorem]{Proposition}

\begin{document}

% If your paper is accepted and the title of your paper is very long,
% the style will print as headings an error message. Use the following
% command to supply a shorter title of your paper so that it can be
% used as headings.
%
%\runningtitle{I use this title instead because the last one was very long}

% If your paper is accepted and the number of authors is large, the
% style will print as headings an error message. Use the following
% command to supply a shorter version of the authors names so that
% they can be used as headings (for example, use only the surnames)
%
%\runningauthor{Surname 1, Surname 2, Surname 3, ...., Surname n}

\twocolumn[

\aistatstitle{Neural SPDE solver for uncertainty quantification \\ in high-dimensional space-time dynamics}

%\aistatsauthor{ Author 1 \And Author 2 \And  Author 3 }

%\aistatsaddress{ Institution 1 \And  Institution 2 \And Institution 3 } ]

\begin{center}
\textbf{Maxime Beauchamp, Hugo Georgenthum and Ronan Fablet} \\ \ \\
IMT Atlantique Bretagne Pays de la Loire  \\ \ \\
\end{center}
]

\begin{abstract}
Historically, the interpolation of large geophysical datasets has been tackled using methods like Optimal Interpolation (OI) or model-based data assimilation schemes. However, the recent connection between Stochastic Partial Differential Equations (SPDE) and Gaussian Markov Random Fields (GMRF) introduced a novel approach to handle large datasets making use of sparse precision matrices in OI. Recent advancements in deep learning also addressed this issue by incorporating data assimilation into neural architectures: it treats the reconstruction task as a joint learning problem involving both prior model and solver as neural networks. Though, it requires further developments to quantify the associated uncertainties. In our work, we leverage SPDE-based Gaussian Processes to estimate complex prior models capable of handling non-stationary covariances in space and time. We develop a specific architecture able to learn both state and SPDE parameters as a neural SPDE solver, while providing the precision-based analytical form of the SPDE sampling. The latter is used as a surrogate model along the data assimilation window. Because the prior is stochastic, we can easily draw samples from it and condition the members by our neural solver, allowing flexible estimation of the posterior distribution based on large ensemble. We demonstrate this framework on realistic Sea Surface Height datasets. Our solution improves the OI baseline, aligns with neural prior while enabling uncertainty quantification and online parameter estimation.
\end{abstract}

\section{Introduction}

Over the last decade, the emergence of large spatio-temporal datasets both coming from remote sensing satellites and model-based numerical simulations has been noticed in Geosciences. As a consequence,  the need for statistical methods able to handle both the size and the underlying physics of these data is growing. Data assimilation (DA) is the traditional framework used by geoscientists to merge these two types of information, data and model, by propagating information from observational data to areas with missing data. Broadly speaking, two main categories of DA \citep{evensen_data_2009, evensen_2022} exist: variational DA and statistical DA. They both aim at minimizing some energy or functional involving a model-based dynamical prior and an observation term. Importantly, statistical DA addresses jointly interpolation and uncertainty quantification issues \citep{tandeo_2020}. Besides, under linear and Gaussian hypotheses, statistical DA stems from the family of the so-called Optimal Interpolation (OI) techniques, see e.g. \citet{traon_1998}. OI is also known as (simple) Kriging in Spatial Statistics \citep{chiles_2012, chiles_2018} and it directly relates to BLUE \citep{asch_data_2016} in the data assimilation formulation, being at the core of the statistical DA methods. The OI implies to factorize dense covariance matrices which turns out to be an issue when the size of the spatio-temporal datasets is large. Reduced-rank approximations, see e.g. \citet{cressie_statistics_2015}, have already been investigated to tackle this specific problem. More recently, the use of sparse covariance matrices has also been proposed by using tapering strategies \citep{furrer_2006, bolin_2016} or by making use of the link seen by \citet{lindgren_2011} between Stochastic Partial Differential Equations (SPDE) and Gaussian Processes. For the latter, if the original link was made through the Poisson SPDE equation \citep{whittle_1953}, it can be extended to more complex linear SPDE involving physical processes such as advection or diffusion processes\citep{lindgren_2011, fuglstad_2015a, fuglstad_2015b, clarotto_2022}. Thus, it opens new avenues to cope with large-scale observational datasets in geosciences while making use of the underlying physics of such processes. Let note that the so-called SPDE-based approach can also be used as a general spatio-temporal model, even if it is not physically motivated, since it provides a flexible way to handle local anisotropies of a large set of geophysical processes. It has known numerous applications in the past few years, see e.g. \citet{Sigrist_2015, Fuglstad_2015}. These applications generally rely on off-line strategies \citep{Fuglstad_2015}. As the parameters of SPDE should vary across space and time for most of the case-studies, this remains a critical shortcoming for the uptake of SPDE formulations in real-world applications.\\

From another point of view, deep learning frameworks provide new means to develop data-driven interpolation schemes. While the missing data rates and sampling patterns encountered in geoscience make less relevant interpolation approaches introduced in computational imaging and computer vision, recent advances bridge deep learning and data assimilation \citep{boudier_2023,fablet2020joint}. These studies leverage neural parameterizations of elementary components of DA schemes and train end-to-end DA solvers from data. Interestingly, in \citet{fablet2020joint}, the so-called 4DVarNet neural scheme exploits a trainable gradient-based solver of a variational DA formulation, which can also involve trainable components. This generic framework also applies to OI \citep{beauchamp_2023b} with a linear scaling of the solution on the number of space-time variables, leading to a significant speed up in the computation of the solution. Real-world applications \cite{oceanbench} of these neural interpolation schemes involve purely data-driven priors and do not address uncertainty quantification \cite{beauchamp_2023a}.

%While the former leverages the automatic differen an iterative gradient-based LSTM solver to minimize a variational cost, close to what is encountered in 4DVar data assimilation \citep{carassi_2018}. In this variational cost, the dynamical prior is no longer model-based but is stated as a trainable neural network learnt during the training process. Then, automatic differentiation is used to compute the gradient of the variational cost during the gradient-based iterations, instead of requiring the computation of complex and costly adjoint models \citep{asch_data_2016}. Drawing from this framework, a neural optimal interpolation scheme has also been proposed \citep{beauchamp_2023b} to reach OI performance with a linear scaling of the solution on the number of space-time variables, leading to a significant speed up in the computation of the solution.\\ 

In this work, we leverage both SPDE priors and trainable neural DA solvers to introduce a neural SPDE solver and address jointly interpolation, uncertainty quantification and SPDE calibration issues. Formally, we state the considered problem as the joint inversion of a state trajectory and of space-time-varying SPDE parameters. For a real case-study, namely ocean altimetry \citep{oceanbench}, we reach similar interpolation performance as when using purely data-driven neural priors while bringing the ability to sample in the posterior distribution. The key contributions are four-fold:
\begin{itemize} 
\item We develop the explicit solver of the considered SPDE prior. It relies on the analytical expression for the SPDE-based precision matrix of any state trajectory, based on a finite-difference discretization of the grid covered by the tensors involved in our neural scheme; 
\item We exploit this SPDE parametrization as surrogate prior model in the proposed variational formulation and leverage a trainable gradient-based  solver to address jointly the interpolation of the state trajectory and the estimation of SPDE parameters from irregularly-sampled observations. The end-to-end training of the solver targets both the expectation of the state given the observations, together with the SPDE parametrization maximizing its likelihood given the true states;
\item The SPDE prior paves the way to uncertainty quantification through the sampling of the prior pdf and the conditioning by the neural gradient-based solver.;
%and entails the possibility of generating huge members in the posterior pdf, after conditioning of the prior samples by our neural variational scheme. We report a real-world application to the interpolation
%of sea surface dynamics from satellite-derived observations.
\item We demonstrate how the proposed framework relates to many ideas commonly shared among generative deep learning models;\end{itemize} 
To make easier the reproduction of our results, an open-source version of our code is available \footnote{To be made available in a final version}.

\section{Background}

\subsection{GP and Optimal Interpolation}

For a m-dimensional Gaussian process $\mathbf{x}$ with mean $\mu$ and covariance $\mathbf{P}$:
\begin{align}
\mathbf{x} \sim \mathcal{G P}\left(\mu(\mathbf{x}), \mathbf{P}\left(\mathbf{x}, \mathbf{x}^{\prime}\right)\right),
\end{align}
and a Gaussian likelihood of partial and noisy observations $\mathbf{y} \in \mathbb{R}^p$:
\begin{align}
\mathbf{y} \mid \mathbf{x} & \sim  \mathcal{N} \left(\mathbf{H}_\Omega \cdot \mathbf{x}_, \mathbf{R} \right), 
\end{align}
the posterior $p(\mathbf{x} \mid \mathbf{y})$, can be computed in closed form, at a computational cost $\mathcal{O}\left(n^{3}\right)$. This is the so-called optimal interpolation that states the reconstruction as the minimization of a variational cost:
\begin{equation}
\label{eq: OI}
\mathbf{x}^\star=\arg \min_\mathbf{x}  \left \|\mathbf{y}-\mathbf{H}_\Omega \cdot \mathbf{x}  \right \|^2_{\mathbf{R}}+ \lambda \left \|\mathbf{x}-\mu \right \|^2_{\mathbf{P}}
\end{equation}
with $\mathbf{H}_\Omega$ denotes the observation matrix to map state $\mathbf{x}$ over domain $\cal{D}$ to the observed domain $\Omega$. $\|\cdot\|^2_{\mathbf{R}}$ is the Mahanalobis norm w.r.t. the covariance of the observation noise $\mathbf{R}$ and $\|\cdot \|^2_\mathbf{P}$ the Mahanalobis distance with prior covariance  $\mathbf{P}$. %The latter decomposes as a 2-by-2 block matrix $[\mathbf{P}_{\Omega,\Omega} \mathbf{P}_{\Omega,\overline{\Omega}};\mathbf{P}_{\overline{\Omega},\overline{\Omega}} \mathbf{P}^{T}_{\Omega,\overline{\Omega}}]$ with $\mathbf{P}_{\cal{A},\cal{A}'}$ the covariance between subdomains $\cal{A}$ and $\cal{A}$ of domain $\cal{D}$.

The OI variational cost (\ref{eq: OI}) being linear quadratic, the solution of the optimal interpolation problem is given by:
\begin{equation}
\label{eq: OI solution}
\mathbf{x}^\star= \mu + \mathbf{K} \cdot \mathbf{y}
\end{equation}
with $\mathbf{K}$ referred to as the Kalman gain $ \mathbf{P} \mathbf{H}\tr_\Omega ( \mathbf{H}_\Omega \mathbf{P} \mathbf{H}\tr_\Omega + \mathbf{R} )^{-1}$ \citep{asch_data_2016}, where $\mathbf{P} \mathbf{H}\tr_\Omega \in \mathbb{R}^{m \times p}$ and $\mathbf{H}_\Omega \mathbf{P} \mathbf{H}\tr_\Omega  \in \mathbb{R}^{p \times p}$ resp. denotes the (grid,obs) and (obs,obs) prior covariance matrix. For high-dimensional  spatio-temporal state sequences of length $N$ and large observation domains, the computation of the Kalman gain becomes rapidly intractable due to the inversion of a $|\Omega|\times|\Omega|$ covariance matrix. This has led to a rich literature to solve minimization \ref{eq: OI} without requiring the above-mentioned $|\Omega|\times|\Omega|$ matrix inversion, among others gradient-based solvers using matrix-vector multiplication (MVMs) reformulation \cite{Pleiss_2020, Aune_2013, Charlier_2020, Cutajar_2016} and methods based on sparse matrix decomposition with tapering \citep{furrer_2006,romary_2018}.

\subsection{Spatio-temporal GP as SDE}

In the last decade, the connection proposed by \citet{lindgren_2011} between Stochastic Partial Differential Equations (SPDE) and Gaussian Markov Random Fields (GMRF) introduced a novel approach to handle large datasets making use of sparse precision-based matrix parameterizations in OI schemes \citep{carrizovergara_2018, clarotto_2022, fuglstad_2015a}.\\

Starting from a general formulation of the continuous state space $\mathbf{x}$, its dynamical evolution states as a stochastic differential equation (SDE):
\begin{align}
\label{gen_sde}
d\mathbf{x} & =\mathcal{F}(\x,t)dt +\mathcal{G}(\x,t) d\mathbf{w}
\end{align}
where $\mathcal{F}({\cdot,t}) : \mathbb{R}^m \rightarrow \mathbb{R}^m$, $\mathcal{G}({\cdot,t}) : \mathbb{R}^m \rightarrow \mathbb{R}^{m \times m}$ and $\mathbf{w}(\mathbf{s}, t)$ is the standard Wiener process (Brownian motion).

\begin{prop}
When operator $\mathcal{F}(\x,t)$ is linear and noise effect $\mathcal{G}(\x,t)$ denotes the square root of the precision matrix $\mathbf{Q}{z,t}$ of the colored noise $z_t=dw/dt$, then denoted $\mathbf{F}_t$ and $\mathbf{L}_{z,t}$, one possible discretization of SDE (\ref{gen_sde}) gives the following state space trajectory $\mathbf{x} = \lbrace \mathbf{x}_0, \cdots, \mathbf{x}_N \rbrace$, which identifies to state equation in data assimilation schemes:
\begin{align}
\label{spde_discrete}
\mathbf{x}_{t+1} & = \mathbf{M}_{t+1}\mathbf{x}_t + \boldsymbol{z}_{t+1}, \ \ \ \ \ \ \ 
\boldsymbol{z}_{t+1} &\sim \mathcal{N}(\mathbf{0}, \mathbf{Q}_{z,t+1}^{-1}) \nonumber \\
\mathbf{x}_{0} & \sim \mathcal{N}(\mathbf{0}, \mathbf{P}_0) 
\end{align}
where $\mathbf{M}_{t+1}$ is the state transition matrix from time $t$ to $t+1$:
\begin{align*}
\mathbf{M}_{t+1} &= \left( \mathbf{I} - dt\mathbf{F}_{t+1} \right)^{-1} \in \mathbb{R}^{m \times m}
\end{align*}
and $\mathbf{T}_{t+1} = \mathbf{M}_{t+1}\mathbf{L}_{z,t+1}  \in \mathbb{R}^{m \times m}$
\end{prop}

\textit{Proof} comes along with the backward Euler discretization of SDE (\ref{gen_sde}):
\begin{align*}
& \frac{\mathbf{x}_{i+1} - \mathbf{x}_{i}}{dt} & = & \mathbf{F}_{i+1} \mathbf{x}_{i+1} + \mathbf{L}_{z,i+1} \mathbf{z}_{i+1} \\
& \left(\mathbf{I} - \mathbf{F}_{i+1}\right) \mathbf{x}_{i+1} & = & \mathbf{x}_{i} + \mathbf{L}_{z,i+1} \mathbf{z}_{i+1}\\
& \ &=& \mathbf{M}_{i+1}\mathbf{x}_i +  \mathbf{T}_{i+1}\boldsymbol{z}_{i+1}
\end{align*}

More often, instead of involving complex colored noise effects $\mathbf{L}_{z,t}$, a simple time-dependent variance regularization $g(t) : \mathbb{R} \rightarrow \mathbb{R}$ is used and $\boldsymbol{z}_{t+1} \sim \mathcal{N}\left(\mathbf{0}, g^2(t) \mathbf{I} \right)$ is pure white noise.

\iffalse
\begin{figure}[h]
\includegraphics[width=8cm]{figures/OI_SPDE}
\caption{SPDE-based OI scheme: the assimilation is not sequential. The inversion scheme makes use of the precision matrix of the state sequence $\lbrace{ \mathbf{x}_0,\cdots, \mathbf{x}_N \rbrace}$ to compute the Gaussian posterior pdf}
\label{oi_spde}
\end{figure}
\fi

The solution of Eq. (\ref{spde_discrete}) is a GP, see e.g. \citet{lindgren_2011, sarkka_solin_2019}. When the likelihood is Gaussian, the posterior is also a GP, then available in closed form. Sequential Kalman recursions are generally used with computational scaling $\mathcal{O}\left(N m^{3}\right)$ to solve efficiently the posterior distribution by a preliminary filtering (forward pass) followed by a smoothing (backward pass). 

\subsection{Sparse precision matrix and OI}

An Optimal Interpolation scheme can also be used, but its time computational complexity is generally prohibitive $\mathcal{O}\left((Nm)^{3}\right)$ \citep{asch_data_2016}. Hopefully, efficient inference algorithms with much lower time computational complexity $\mathcal{O}\left((Nm)^{3/2}\right)$ \citep{lindgren_2011, chiles_2018} can be used by switching from dense covariance matrices $\bP$ to sparse precision matrix $\mathbf{Q}$, see Proposition 2.% and Figure \ref{oi_spde}.

\begin{figure*}
\begin{prop}
Because of the formulation of $\mathbf{M}_t$ and $\mathbf{T}_t$, the precision matrix $\mathbf{Q} = \mathbf{P}^{-1}$ writes:
\begin{align}
\hspace{-1.5cm}
\label{global_prec_mat_2}
\mathbf{Q} &= \frac{1}{dt} \begin{bsmallmatrix} 
\mathbf{P}_0^{-1}+\mathbf{Q}_{z,1}  & -\mathbf{Q}_{z,1}\mathbf{M}_1^{-1} & 0 & 0 & 0 &\dots& 0 \\
-\left(\mathbf{M}_1\tr\right)^{-1}\mathbf{Q}_{z,1} & \mathbf{M}\tr_1\mathbf{Q}_{z,1}\mathbf{M}_1 +\mathbf{Q}_{z,2} & -\mathbf{Q}_{z,2}\mathbf{M}_2^{-1} &0  & 0 & \dots &0 \\
0 & -\left(\mathbf{M}_2\tr\right)^{-1}\mathbf{Q}_{z,2} & \mathbf{M}\tr_2\mathbf{Q}_{z,2}\mathbf{M}_2 + \mathbf{Q}_{z,3} & -\mathbf{Q}_{z,3}\mathbf{M}_3^{-1} & 0 & \dots & 0\\
0 &\ddots &\ddots &\ddots & \ddots & \ddots & 0 \\
0 &\ddots &\ddots &\ddots & \ddots & \ddots & 0 \\
\vdots &\ddots &\ddots &\ddots & -\left(\mathbf{M}_{N-1}\tr\right)^{-1}\mathbf{Q}_{z,N-1} & \mathbf{M}\tr_{N}\mathbf{Q}_{z,N-1} \mathbf{M}_{N-1} + \mathbf{Q}_{z,N} & -\mathbf{Q}_{z,N}\mathbf{M}_{N}^{-1} \\
0 &\ddots &\ddots &\ddots & 0 & -\left(\mathbf{M}_{N}\tr\right)^{-1}\mathbf{Q}_{z,N} & \mathbf{M}\tr_{N}\mathbf{Q}_{z,N}\mathbf{M}_{N} \\
\end{bsmallmatrix}
\end{align}
\end{prop}

\textit{Proof} is given in App. 1.
\end{figure*}

When considering the augmented state vector $[\y,\x]$ with covariance matrix $\bP_G$:
\begin{align}
\bP_G &=  \begin{bmatrix} \bP_{\mathbf{y}\mathbf{y}} &  \bP_{\mathbf{y}\mathbf{x}} \\ \bP_{\mathbf{x}\mathbf{y}}  & \bP_{\mathbf{x}\mathbf{x}} \end{bmatrix}  &=&
\begin{bmatrix} \bP & \bP \mathbf{H}_\Omega\tr \\
 \mathbf{H}_\Omega \bP & \mathbf{H}_\Omega \bP \mathbf{H}_\Omega\tr + \mathbf{R}
\end{bmatrix}
\end{align}
Then, the precision matrix $\mathbf{Q}_G$ is still sparse and writes:
\begin{align}
\mathbf{Q}_G &=  \begin{bmatrix} \mathbf{Q}_{\mathbf{y}\mathbf{y}} &  \mathbf{Q}_{\mathbf{y}\mathbf{x}} \\ \mathbf{Q}_{\mathbf{x}\mathbf{y}}  & \mathbf{Q}_{\mathbf{x}\mathbf{x}} \end{bmatrix}
&=&\begin{bmatrix} \mathbf{Q} +  \mathbf{H}_\Omega\tr \mathbf{R}^{-1} \mathbf{H}_\Omega & -\mathbf{H}_\Omega\tr\mathbf{R}^{-1}  \\  -\mathbf{R}^{-1}\mathbf{H}_\Omega  & \mathbf{R}^{-1},
\end{bmatrix}
\end{align}
and by noticing that:
\begin{subequations}
\begin{align}
&  \bP_G\mathbf{Q}_G & = &  \begin{bmatrix} \mathbf{I}_{\mathbf{y}\mathbf{y}} &  \mathbf{0}_{\mathbf{y}\mathbf{x}} \\ \mathbf{0}_{\mathbf{x}\mathbf{y}}  & \mathbf{I}_{\mathbf{x}\mathbf{x}} \end{bmatrix}\\
\label{id1}
& \bP_{\mathbf{y}\mathbf{y}}\mathbf{Q}_{\mathbf{y}\mathbf{x}} + \bP_{\mathbf{y}\mathbf{x}}\mathbf{Q}_{\mathbf{x}\mathbf{x}} & = &\mathbf{0}_{\mathbf{y}\mathbf{x}} \\
\label{id2}
 &\mathbf{Q}_{\mathbf{y}\mathbf{x}} \mathbf{Q}_{\mathbf{x}\mathbf{x}}^{-1} + \bP_{\mathbf{y}\mathbf{y}}^{-1}\bP_{\mathbf{y}\mathbf{x}} & = & \mathbf{0}_{\mathbf{y}\mathbf{x}}, 
\end{align}
\end{subequations}
adding Eqs. (\ref{id1}) and (\ref{id2}) and finally transposing, we rewrite the Kalman gain $\mathbf{K} \in \mathbb{R}^{m \times p}$ in terms of precision matrix:
\begin{align}
\label{equiv}
 \mathbf{K} &= \bP_{\mathbf{x}\mathbf{y}}\bP_{\mathbf{y}\mathbf{y}}^{-1} \nonumber \\
             &= -\mathbf{Q}_{\mathbf{x}\mathbf{x}}^{-1} \mathbf{Q}_{\mathbf{x}\mathbf{y}} &=&  \left( \mathbf{Q} +  \mathbf{H}_\Omega\tr \mathbf{R}^{-1} \mathbf{H}_\Omega \right)^{-1}\mathbf{H}_\Omega\tr\mathbf{R}^{-1}
\end{align}
and the posterior precision matrix is $\mathbf{Q}(\x \mid \y)=\mathbf{Q}(\x) +  \mathbf{H}_\Omega\tr \mathbf{R}^{-1} \mathbf{H}_\Omega$

%\section{Joint learning of posterior and SPDE prior parametrization}
\section{Neural solver for posterior and SPDE prior parametrization}
%\section{Proposed approach}

For high-dimensional state space, even sparse precision-based inference may lead to high complexity. In that case, matrix-free approaches can be used using iterative solvers without explicitly build and store the precision matrix $\mathbf{Q}$ \citep{saad_2003, pereira_2019, pereira_2022}. Still, all these methods became hardly tractable when the problem of estimating the posterior distribution $p(\x \mid y)$ comes jointly with the estimation of the prior SPDE parameter $\Theta$. For stationary SPDEs, \citet{clarotto_2022} use the Broyden, Fletcher, Goldfarb, and Shanno optimization algorithm \citep{NoceWrig06} involving the second-order derivative of the objective function, while for pure spatial GPs, \cite{fuglstad_2015a} forms a hierarchical spatial model for $p(\x \mid y,\Theta)$ whose inference is handled numerically by the INLA methodology \citep{rue_2017}. All these methods are either time-consuming or comes with simplified assumptions on the SPDE formulation. In this Section, we show how we can escape from solving the linear system and draw from the equivalent variational OI scheme to learn jointly how to compute the posterior and estimate the SPDE parameters. Also, if the training dataset is large enough, our approach provides an efficient way to estimate \textit{online} the SPDE parametrization for any new sequence of input observations, without any additional inference to make.

\subsection{Neural variational scheme}

Variational formulations have also been widely explored to solve inverse problems. Similarly to (\ref{eq: OI}), the general formulation involves the sum of a data fidelity term and of a prior term \citep{asch_data_2016}. In a model-driven approach, the latter derives from the governing equations of the considered processes. For instance, data assimilation in geoscience generally exploits PDE-based terms to state the prior on some hidden dynamics from observations. In signal processing and computational imaging, similar formulations cover a wide range of inverse problems, including inpainting issues \citep{bertalmio_navier-stokes_2001}:
\begin{equation}
\label{eq: proj-based OI}
\mathbf{x}^\star=\arg \min_\mathbf{x} \left \|\mathbf{y}-\mathbf{H}_\Omega \cdot \mathbf{x}\right \|^2+ \lambda \left \| \mathbf{x}-\Phi (\mathbf{x}) \right \|^2  
\end{equation}
where $\lambda$ a positive scalar to balance the data fidelity term and the prior regularization term. Prior $\Phi$ can be seen as a projection operator, comprising gradient-based formulations using finite-difference approximations, proximal operators as well as plug-and-play priors \cite{aubert_mathematical_2006,lucas_using_2018,mccann_convolutional_2017}. As mentioned above, these formulations have also gained interest for the definition of deep learning schemes based on the unrolling of minimization algorithms \citep{andrychowicz2016learning,aggarwal_modl_2019} for Eq. (\ref{eq: proj-based OI}). \\

We benefit from automatic differentiation tools associated with neural operators to investigate iterative gradient-based solvers of Eq. (\ref{eq: proj-based OI}), denoted as $\Gamma$ in the following. To speed up the gradient descent, we rely on a neural operator ${\mathcal{K}}$ that combines an LSTM cell \citep{Shi_2015} together with a final linear layer to map the hidden state of the LSTM cell to the space spanned by state $\mathbf{x}$. The state updates writes:

\begin{align}
 \mathbf{x}^{(k+1)} = \mathbf{x}^{(k)} - \mathcal{K}\left[ \nabla_\mathbf{x} \mathcal{J}_\Phi\left ( \mathbf{x}^{(k)} , \mathbf{y}, \Omega \right) \right ]
\end{align}

Overall, the neural architecture is denoted $\Psi_{\Gamma} (\mathbf{x}^{(0)},\mathbf{y},\Omega)$  given some initial conditions $\mathbf{x}^{(0)}$ and observations $\mathbf{y}$ on domain $\Omega$.

\subsection{SPDE as interpretable prior} 

Here, we use the neural scheme introduced above as neural solver for an SPDE-based formulation of the interpolation problem. 

Prior operator $\Phi$ is not easily interpretable, simply acting as a projection of state $\mathbf{x}$ to help in the gradient-based minimization process. Here, we bring both interpretability and stochasticity in the neural scheme by considering as prior surrogate model a linear SPDE (\ref{spde_discrete}). Operator $\mathbf{F}$ states as the finite difference discretization scheme ($\mathit{FDM}$) of a fractional advection-diffusion operator:
\begin{align*}
\mathbf{F} = \mathit{FDM}\left(\left\lbrace \boldsymbol{\kappa}^2_t   - \nabla \cdot\mathbf{m}_t  - \nabla \cdot\mathbf{H}_t \nabla \right \rbrace^{\alpha/2}\right)
\end{align*}
with $\mathbf{m}_t$ and $\mathbf{H}_t$ resp. the advection vector and diffusion tensor, $\boldsymbol{\kappa}_t$ acts as a scaling parameter and $\alpha$ relates to the smoothness of the underlying GP.  Operator $\mathbf{G}$ is  simplified so that the right-hand side of the SPDE is a white noise with variance regularization $\boldsymbol{\tau}_t$. Back to the OI variational problem (\ref{eq: OI}) and the results obtained for precision matrix $\mathbf{Q}$ of state sequences $\mathbf{x}=\lbrace{ \mathbf{x}_0, \cdots, \mathbf{x}_T \rbrace}$, see Proposition 2, the variational cost to minimize becomes:
\begin{align}
\label{eq: proj-based OI new}
\mathbf{x}^\star=\arg \min_\mathbf{x} \left \|\mathbf{y}-\mathbf{H}_\Omega \cdot \mathbf{x}\right \|^2+ \lambda \cdot \mathbf{x}\tr \mathbf{Q} \mathbf{x} 
\end{align}
which is a specific case of Eq. (\ref{eq: proj-based OI}) where $||\x -\Phi(\x) ||^2=\mathbf{x}\tr \mathbf{Q} \mathbf{x}$ with $\Phi=(1-\bS)$, and $\bS$ is the square root of the precision matrix $\mathbf{Q}$.

\subsection{Training scheme}
Finding the best set of parameters for the SPDE to optimize the reconstruction and helps to uncertainty quantification may be intricate. Drawing from our neural variational framework, the trainable prior is now SPDE-based, and the parameters $\boldsymbol{\Theta} = \begin{bmatrix} \boldsymbol{\kappa} & \mathbf{m} & \mathbf{H} & \boldsymbol{\tau} \end{bmatrix}^{\mathrm{T}}$ are now embedded in an augmented state formalism, i.e.:
\begin{eqnarray}
\label{eq: augmented formulation}
 \tilde{\mathbf{x}} = \begin{bmatrix} \mathbf{x} & \boldsymbol{\Theta} \end{bmatrix}^{\mathrm{T}}
\end{eqnarray}
Latent parameter $\boldsymbol{\Theta}$ is non stationary in both space and time and its size directly relates to the size $N$ of the state sequence. The joint learning of SPDE parametrization $\boldsymbol{\Theta}$ and solvers $\Gamma$ for the reconstruction $\mathbf{x}^\star$ states as the minimization of:
\begin{equation}
\label{eq: E2E loss}
   \arg \min_{\boldsymbol{\Theta},\Gamma} \mathcal{L}(\mathbf{x},\boldsymbol{\Theta}^\star,\mathbf{x}^\star) \mbox{  s.t.  } 
   \tilde{\mathbf{x}}^\star = \Psi_{\Gamma}  (\tilde{\mathbf{x}}^{(0)},\mathbf{y},\Omega)
\end{equation}
with $\mathcal{L}(\mathbf{x},\boldsymbol{\Theta}^\star,\mathbf{x}^\star)=\lambda_1\mathcal{L}_1(\mathbf{x},\mathbf{x}^\star)+\lambda_2\mathcal{L}_2(\mathbf{x},\boldsymbol{\Theta}^\star)$. $\mathcal{L}_1(\mathbf{x},\mathbf{x}^\star)=||\mathbf{x}-\mathbf{x}^\star||^2$ is the reconstruction cost, i.e. the MSE w.r.t Ground Truth and $\mathcal{L}_2(\mathbf{x},\Theta^\star)$ is the negative log-likelihood $-p(\mathbf{x} \mid \Theta^\star)$ with:
\begin{align}
p(\mathbf{x} \mid \Theta^\star)= -|\mathbf{Q}(\boldsymbol{\Theta}^\star)| + \mathbf{x}^{\mathrm{T}} \mathbf{Q}(\boldsymbol{\Theta}^\star)\mathbf{x}
\end{align}
used as the prior regularization cost. The log-determinant of the precision matrix $\log|\mathbf{Q}(\boldsymbol{\Theta}^\star)|$ is usually difficult to handle. 

\begin{prop} 
Based on its particular block-sparse structure and the notations already introduced for the spatio-temporal precision matrix $\mathbf{Q}$: 
\begin{align}
\label{log_det_Q}
\log|\mathbf{Q}(\Theta^\star)| & = \log|\mathbf{P}_0^{-1}| + 2 \sum_{i=1}^L \sum_{j=1}^m \log \mathbf{L}_i (j,j)
\end{align}       
where $\mathbf{L}_i$ denotes here the Cholesky decomposition of $\mathbf{S}_k^{-1}$ and $\mathbf{S}_k=\mathbf{T}_k\mathbf{T}_k\tr$.\\ 
\end{prop}
Based on \citet{powell_2011} to compute the determinant of an $N \times N$ complex block matrix in terms of its constituent blocks, \textit{Proof} is given in App. 2. Note that $|\mathbf{S}_k^{-1}|$ is now the determinant of a sparse matrix $\in \mathbb{R}^m$, symmetric and positive deﬁnite matrix. The computation of its determinant can be obtained by Cholesky decomposition. Also, because precision matrix $\mathbf{Q}$ and submatrices $\mathbf{S}_k$ are used in the forward pass of our architecture, the backward pass for their Cholesky decomposition is needed. Given a symmetric, positive deﬁnite matrix $\mathbf{A}$, its
Cholesky factor $\mathbf{L}$ is lower triangular with positive diagonal, such that the forward pass of the Cholesky decomposition is defined as $\mathbf{A}=\mathbf{L}\mathbf{L}\tr$. Given the output gradient $\overline{\mathbf{L}}$ and the Cholesky factor $\mathbf{L}$, the backward pass compute the input gradient $\overline{\mathbf{A}}$ defined as :
\begin{align}
\overline{\mathbf{A}} = \frac{1}{2} \mathbf{L}^{-\mathrm{T}} ltu(\mathbf{L}\tr\overline{\mathbf{L}})\mathbf{L}^{-1}
\end{align}
where $ltu(\cdot)$ generates a symmetric matrix by copying the lower triangle to the upper
triangle. 

\subsection{Sampling of the posterior}

Inspired by kriging-based conditioning \citep{wackernagel_2003}, we draw SPDE-based conditional simulations with our neural scheme:
\begin{align}
  \label{simu_cond}
 \mathbf{x}^{\star,i}  = \mathbf{x}^\star + (\mathbf{x}_{s}^i - \mathbf{x}_s^{\star,i})
\end{align}
where $\mathbf{x}^\star$ denotes the neural-based interpolation, $\mathbf{x}^i_{s}$ is an SPDE  simulation of the space-time trajectory $\lbrace{ \mathbf{x}_0, \cdots, \mathbf{x}_N \rbrace}$ based on the parameters $\Theta^\star$ and $\mathbf{x}^{\star,i}_{s}$ is the neural reconstruction of this non-conditional simulation, using as pseudo-observations a subsampling of $\mathbf{x}^i_{s}$ based on the actual data locations.\\

\section{Generative models as related works}
\label{rel_works}

The targeted interpolation problem, stated as the sampling of the posterior pdf of a state given some partial observations, relates to conditional generative models, where  the conditioning results from the partial observations. Generative models have received a large attention in the deep learning literature with a variety of neural schemes including among others GANs \citep{goodfellow_2014}, VAEs \citep{Kingma_2022}, normalizing flows \citep{Dinh_2017} and diffusion models \citep{Ho_2020}. We discuss further these connections below.

As we rely on an explicit SPDE-based state-space formulation, we can make explicit the gradient of the inner variational cost $\mathcal{J}(\mathbf{x},\mathbf{y},\Omega)$. From Eq. (\ref{eq: proj-based OI new}), considering a covariance matrix $\R$ for the observation noise, we can derive the following expression:
%Our approach also shares many ideas with so-called DL-based generative models. In this work, we proposed a joint learning of the reconstruction together with the parametrization of a stochastic prior. As detailed in Eq. (\ref{eq: E2E loss}), this implies the use of a reconstruction loss and a regularizer, the negative log-likelihood (NLL) of the SPDE parameters. When involving a linear operator $\Phi$ together with observation and model (prior) error noise matrices $\R$ and $\Q$, the gradient of the inner variational cost $\mathcal{J}(\mathbf{x},\mathbf{y},\Omega)$ writes:
\begin{align}
  \label{equiv_grad4Dvar_lgv}
  \nabla_{\mathbf{x}} \mathcal{J}(\mathbf{x},\mathbf{y},\Omega) & = \nabla_{\mathbf{x}} \left [\mathbf{d}\tr\mathbf{R}^{-1}\mathbf{d} + \mathbf{x}\tr\mathbf{Q}^{-1}\mathbf{x} \right] \nonumber  \\
  & = -\mathbf{H}\tr_\Omega\mathbf{R}^{-1}\mathbf{d} +\mathbf{Q}^{-1}\mathbf{x} \nonumber \\
  & = \nabla_{\mathbf{x}} \mathrm{log} p(\mathbf{y}|x) + \nabla_{\mathbf{x}} \mathrm{log} p(\mathbf{x})
\end{align} 
where $\mathbf{d}=\mathbf{y}-\mathbf{H}_\Omega\mathbf{x}$. Score-based approaches \citep{Sohldickstein_2015} exploits similar formulations. However, they directly parameterize the gradient of the likelihoods rather than the likelihoods themselves. Here, we exploit the latter through an observation operator and the SPDE prior. Another important difference with score-based approaches lies in the considered gradient-based procedure to sample the posterior. Score-based schemes generally rely on Langevin dynamics \citep{grenander_1994} {\em i.e.} a stochastic gradient descent for posterior $\mathbf{x}|\mathbf{y}$. Here, we exploit a gradient-based LSTM solver similarly to trainable optimizers exploited in meta-learning schemes \citep{andrychowicz2016learning}. This greatly speeds up the convergence of the inner minimization and allows us to train the overall neural scheme end-to-end. As we do not exploit a stochastic solver, our ability to sample the posterior does not derive from the implementation of Langevin dynamics but we exploit the analytic form of the SPDE to sample in the prior. Future works could explore further whether Langevin dynamics fits within the proposed framework.

%This may be seen as a specific formulation for Langevin dynamics used to draw sample from $p(\mathbf{x}|\mathbf{y})$, except that we introduce the LSTM in Eq. (\ref{equiv_grad4Dvar_lgv}), see also Fig.\ref{gen_mod}a) to speed up the convergence and no noise is introduced along the iterations. This is why our solver $\Gamma$ is deterministic and not stochastic. For now, we only estimate the expectation of $\mathbf{x,\Theta}|\mathbf{y}$ but future works may draw from Langevin dynamics to propose stochastic solvers in our neural variational scheme. \\
\begin{figure}[h]
\subfloat[][Langevin-related dynamics]{\includegraphics[width=8cm]{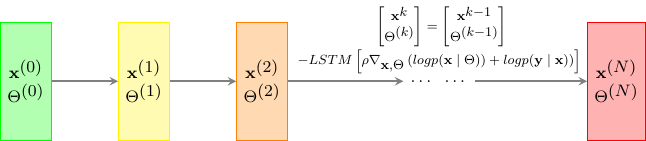}}\\
\subfloat[][SPDE-based VAE formulation]{\includegraphics[width=8cm]{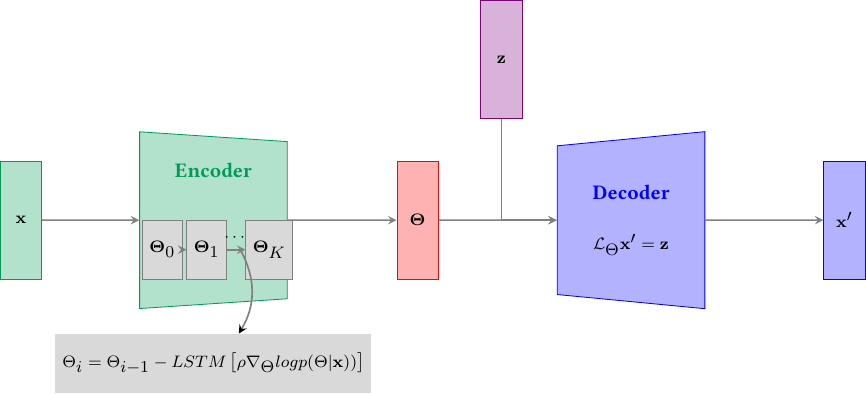}}\\
\subfloat[][Diffusion-based analogy]{\includegraphics[width=8cm]{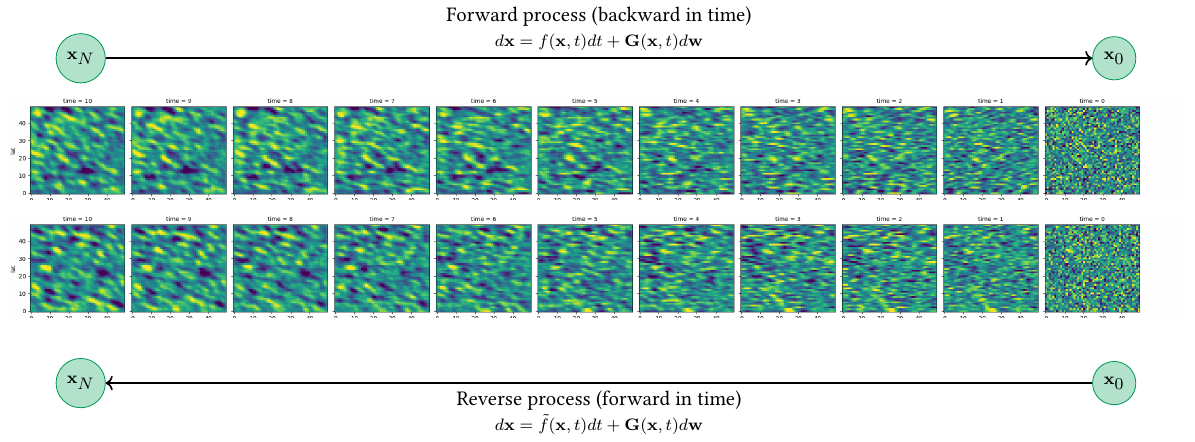}}
\caption{Example of analogies between our combination of SPDE prior and neural solver with generative models}
\label{gen_mod}
\end{figure}

In (\ref{equiv_grad4Dvar_lgv}), we could consider using only the regularization cost if we are only interested to learn SPDE surrogate generative models using as training data gap-free states. In such a case, the proposed formulation relates to variational autoencoder (VAE) formulations \citep{Kingma_2022}, see also Fig.\ref{gen_mod}b), in which the encoder projects the state $\mathbf{x}$ to the parameter space $\Theta$. The encoder is trained on the NLL of $\Theta$. The decoder is simply the SPDE, and does not require any training. Back to the encoder part, it may compress the original high-dimensional input space into a lower-dimensional space, if the SPDE is stationary for instance. More generally, when complex spatio-temporal anisotropies are involved, and without a parametrization model for the latent variable $\Theta$, the latter has high dimensionality (same as the original data or higher), as it is done in diffusion-based models, see e.g. \citep{Sohldickstein_2015, Ho_2020}. Along this line, starting from any initial conditions $\Theta_0$, the optimal set of parameters $\Theta^\star$ (given $\mathbf{x}$) would be obtained by the iterations:
\begin{align*}
  \Theta_i = \Theta_{i+1} - \mathcal{K} \big[ \rho \nabla_\Theta \mathrm{log}\  p(\mathbf{x} \mid \Theta) \big]
\end{align*}
because the target cost function would not be the 4DVar cost anymore but its regularization part (prior cost), i.e. $\mathrm{log}\ p(\mathbf{x} \mid \Theta)$. As already said above, this relates to Langevin dynamics formalism.\\

Eq.(\ref{spde_discrete}) provides a simple way to generate SPDE-driven GP simulations starting from white or colored noise $\mathbf{z}_0$.% to a sequence of spatio-temporal states $\lbrace \mathbf{x}_{0:T} \rbrace$. 
It can be seen as the so-called reverse process in diffusion-based models \citep{Ho_2020}. Because we deal with space-time processes, this reverse process would actually go forward in time, while the corresponding forward process would go backward in time till the initial noise used in the SPDE to generate realistic space-time sequences. Considering our formulation, the reverse process can be reformulated as follows:
\begin{align}
\label{new_spde}
\mathbf{x}_{i+1} & = \mathbf{M}_{i+1}\mathbf{x}_i +  \mathbf{T}_{i+1}\boldsymbol{z}_{i} \nonumber \\
& = \mathbf{x}_i -\mathbf{F}_{i+1}\mathbf{M}_{i+1} \mathbf{x}_i + \mathbf{T}_{i+1}\boldsymbol{z}_{i} 
\end{align} 
because of the Woodbury  matrix identity $\mathbf{M}_{i+1}=\mathbf{I}-\mathbf{F}\mathbf{M}_{i+1}$. By \citep{Anderson_1982}, we also know that given a forward process (data to noise):
\begin{align*}
\mathbf{x}_{i} = \mathbf{x}_{i+1} + \mathbf{f}_{i+1}(\mathbf{x}_{i+1}) + \mathbf{G}_{i+1}\mathbf{z}_{i+1}
\end{align*}
the corresponding reverse process (noise to data) writes:
\begin{align*}
\mathbf{x}_{i+1} = \mathbf{x}_{i} + \mathbf{f}_{i}(\mathbf{x}_{i}) &- \frac{1}{2}\nabla \cdot \left[\mathbf{G}_{i} \mathbf{G}_{i}\tr \right] \\ & - \frac{1}{2}\mathbf{G}_{i} \mathbf{G}_{i}\tr\nabla \mathrm{log} p_i(\mathbf{x}_{i}) +  \mathbf{G}_{i}\boldsymbol{z}_{i}
\end{align*} 
Then, by simple identification, the drift term $\mathbf{f}_{i}(\mathbf{x}_{i})$ is:
\begin{align}
\label{sde_id}
\mathbf{f}_{i}(\mathbf{x}_{i}) = -\mathbf{F}_{i+1}\mathbf{M}_{i+1} & + \frac{1}{2}\nabla \cdot \left[\mathbf{T}_{i+1} \mathbf{T}_{i}\tr \right]   \nonumber \\ &+ \frac{1}{2}\mathbf{T}_{i+1} \mathbf{T}_{i+1}\tr \mathbf{P}_{i+1}^{-1} \mathbf{x}_{i+1}
\end{align} 
since $\nabla \mathrm{log} p_{i+1}(\mathbf{x}_{i+1})= -\mathbf{P}_{i+1}^{-1} \mathbf{x}_{i+1}$
Fig.\ref{gen_mod}c) demonstrates this link when simulating with Eq. (\ref{new_spde}) a GP with global anisotropy on a uniform Cartesian grid with $dx$, $dy$ and $dt$ all set to one and shows how we can retrieve the forward process from Eq. (\ref{sde_id}). This opens a new challenge for future work to estimate the underlying stochastic differential equation of the forward process for space-time sequences, then being able to generate realistic space-time dynamics.

\section{Experiments}

We apply the proposed neural SPDE scheme to a real-world dataset, namely the interpolaton of sea surface height (SSH) fields from irregularly-sampled satellite altimetry observations \cite{oceanbench}. The SSH relates to sea surface dynamics \citep{LeGuillou_2020} and satellite altimetry data are characterized by an average missing data rate above 90\%.

\begin{figure}[h!]
\centering
\vspace{.3in}
\includegraphics[width=7.5cm, trim= 2cm 0 2cm 2cm, clip]{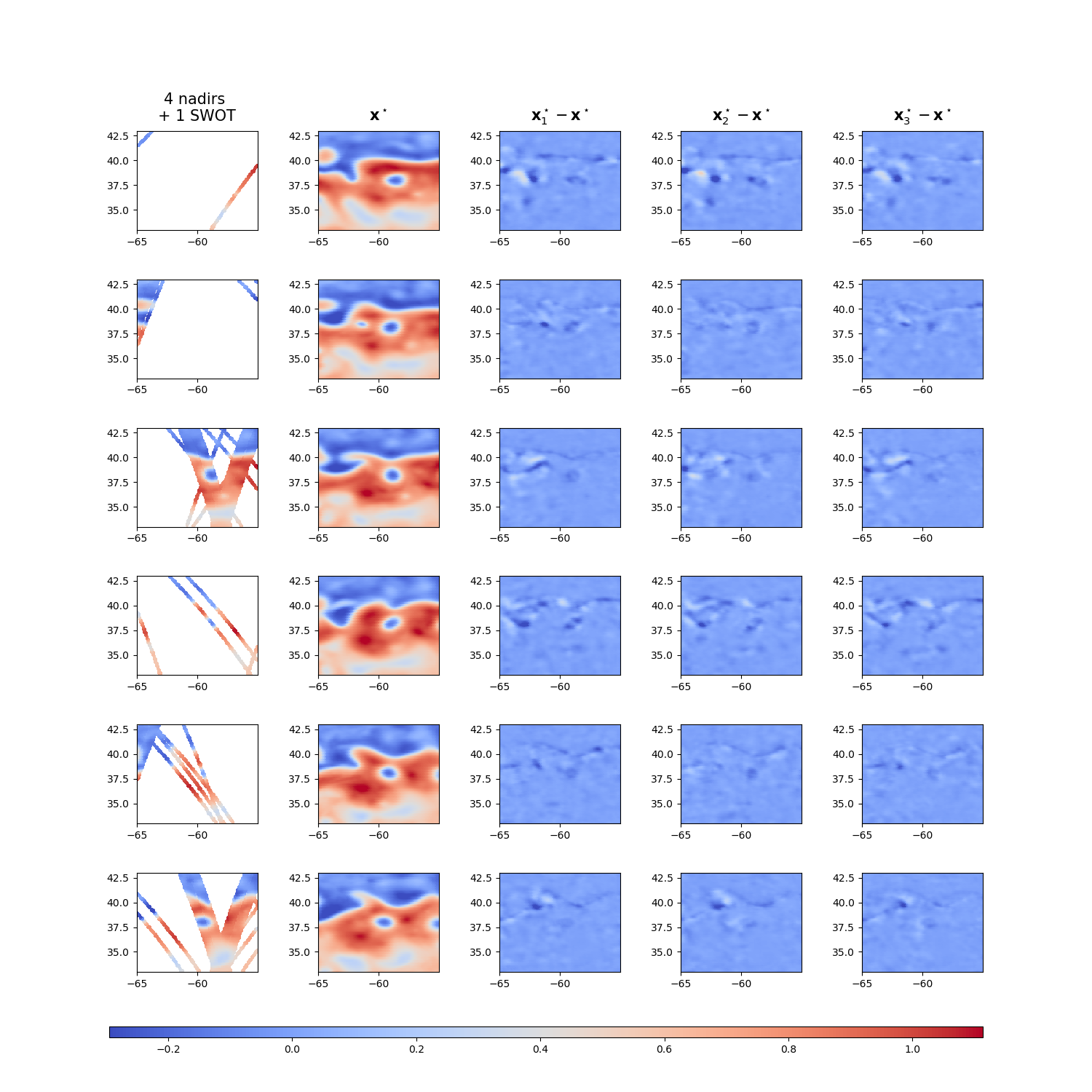}
%\vspace{.3in}
\caption{From left to right: Pseudo-observations (4 nadirs and SWOT), the reconstruction obtained with our neural SPDE variational solver, and deviations to the optimal reconstruction obtained from 3 conditional simulations. Results are given at the center of the data assimilation window  along the 42 days test period and every 6 days (from top to bottom).}
\label{dev_mean}
\end{figure}

\begin{table*}[t!]
\begin{center}
\scalebox{.8}{
\begin{tabular}{l r r r r}
\hline
 & $\mu$(RMSE) & $\sigma$(RMSE) & $\lambda_x$ (degree) & $\lambda_t$ (days) \\
\hline
OI (1 swot + 4 nadirs) & 0.92 & 0.02 & 1.22 & 11.06 \\
BFN & 0.92 & 0.02 & 1.23 & 10.82 \\
DYMOST & 0.91 & 0.02 & 1.36 & 11.91\\
MIOST & 0.93 & 0.01 & 1.35 &10.41  \\
\hline
UNet & 0.92 & 0.02 & 1.25  & 11.33  \\
UNet (time-dependent) & 0.91 & 0.02 & 1.29  & 10.84  \\
4DVarNet - UNet prior  & 0.94 & 0.01 & 1.17 & 6.86 \\
4DVarNet - BilinRes prior  & 0.97 & 0.01 & 0.89 & 4.40 \\
4DVarNet - SPDE prior  & 0.96 & 0.01 & 0.90 & 5.03 \\
\hline
\end{tabular}
}
\end{center}
\caption{Interpolation performance for the satellite altimetry case-study: for each benchmarked models, we report the considered performance metrics averaged on the test period.}
\label{table_scores}
\end{table*}

\paragraph{Experimental setting} We exploit the experimental setting defined in \citep{oceanbench} \footnote{SSH Mapping Data Challenge 2020a: \url{https://github.com/ocean-data-challenges/2020a_SSH_mapping_NATL60}}. It relies on a groundtruthed dataset given by the simulation of realistic satellite altimetry observations from numerical ocean simulations. Overall, this dataset refers to 2{\sc d}+t states for a $10^\circ \times 10^\circ$ domain with 1/20$^\circ$ resolution corresponding to a small area in the Western part of the Gulf Stream. Regarding the evaluation framework, we refer the reader to SSH mapping data challenge above mentioned for a detailed presentation of the datasets and evaluation metrics. The latter comprise the MSE w.r.t the Ground Truth, the minimal spatial and temporal scales resolved. For learning-based approaches, the training dataset spans from mid-February 2013 to October 2013, while the validation period refers to January 2013. In all the reported experiments, we use Adam optimizer over 200 epochs and NVIDIA A100 Tensor Core mono-GPU architectures. All methods are tested on the test period from October 22, 2012 to December 2, 2012. 

\begin{figure}[h]
%\vspace{.3in}
\centering
\includegraphics[width=7.5cm, trim= 2cm 0 2cm 4cm, clip]{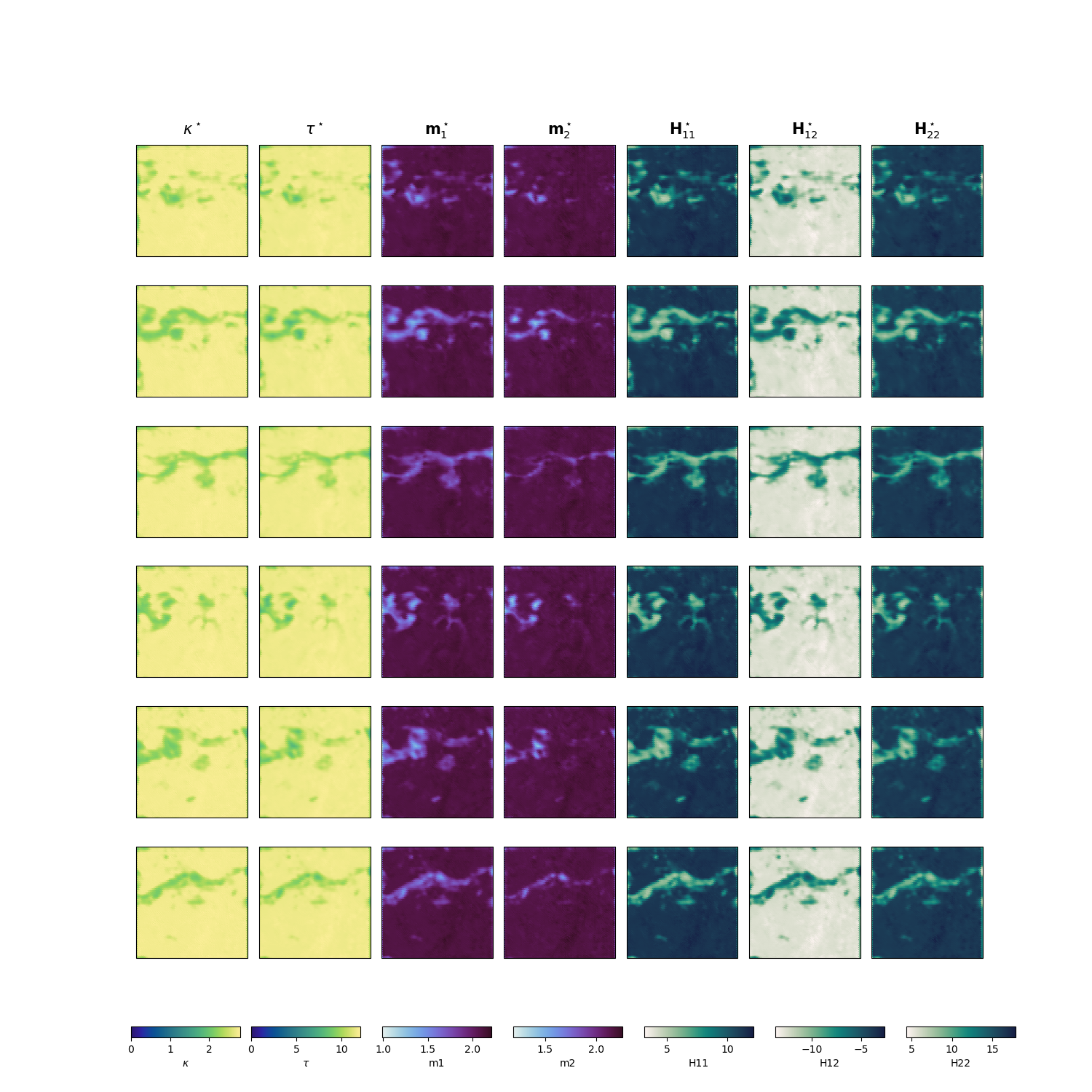}
%\vspace{.3in}
\caption{Parameter estimation of the SPDE prior in the neural scheme along the 42 days test period and every 6 days. From left to right: dumpin and variance parameters, advection fields and diffusion tensors.}
\label{param_osse}
\end{figure}

\vspace{-.3cm}
\paragraph{Benchmarked models}  For benchmarking purposes, we consider the approaches reported in 
\citet{LeGuillou_2020}, namely: the operational baseline (DUACS) based on an optimal interpolation, multi-scale OI scheme MIOST \citep{Ardhuin_2020} and model-driven interpolation schems BFN \citep{LeGuillou_2020} and DYMOST \citep{Ubelmann_2016, Ballarotta_2020}. We also include a state-of-the-art UNet architecture to train a direct inversion scheme \cite{cicek_3d_2016}. As stated in Section \ref{rel_works}, we also implemented a time-dependent UNet to approximate the gradient of the log-prior distribution $\nabla \mathrm{log} p(\x)$ to replace our LSTM solver. Last, for all neural schemes, we consider $N=$ 15 days space-time sequences to account for time scales considered in state-of-the-art OI schemes. Regarding the parameterization of our framework, we consider a bilinear residual architecture for prior $\Phi$, the same UNet used in the direct inversion and the SPDE prior proposed throughout the paper. For the solver $\Gamma$, we use a 2{\sc d} convolutional LSTM cell with 150-dimensional hidden states.

\begin{figure}[h!]
\centering
\includegraphics[width=4cm, trim= .5cm 0 .5cm 0, clip]{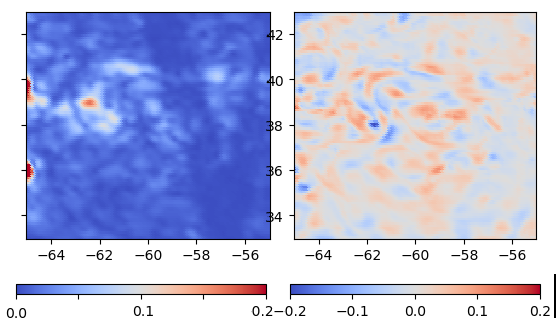}
\caption{Ensemble-based posterior standard deviations and reconstruction error $\mathbf{x} - \mathbf{x}^\star$}
\label{var_errors}
\end{figure}

\vspace{-.3cm}
\paragraph{Results} Table \ref{table_scores} further highlights the performance gain of the proposed scheme. The relative gain is greater than 50\% compared to the operational satellite altimetry processing. We outperform by more than 20\% in terms of relative gain to the baseline MIOST and UNet schemes, which are the second best interpolation schemes. Interestingly, our scheme is the only one to retrieve time scales below 10 days. Figure \ref{dev_mean} displays the reconstruction obtained with our neural variational scheme as well as the deviations to the optimal reconstruction obtained from 3 conditional simulations, see Eq. (\ref{simu_cond}): we can see how their variance is high along the main meander of the Gulf Stream and highly energetic eddies but lower elsewhere. Regarding Figure \ref{param_osse} and the SPDE parameters estimated along the 42 days test period every 6 days, they seem consistent with state $\mathbf{x}$ that partially encodes the SPDE parametrization. The capability of our approach to span parameter distribution that are not standardized is demonstrated, which is particularly the case for the diffusion tensor $\mathbf{H}$. %Playing with both damping and variance regularization parameters provide a flexible way to handle complex GP priors with both low and high marginal variances. 
Last, Figure \ref{var_errors} shows the posterior standard deviations computed when using the SPDE-based generation of 250 members and the reconstruction error $\mathbf{x}-\mathbf{x}^\star$ for the tenth day of the test period. Contrary to OI, the posterior variance is not only conditioned by the observations, it is more continuous and flow dependent. The correlation with the reconstruction error is good, thus it is consistent: when low, the average reconstruction is generally very good.\

\vspace{-.3cm}

\section{Conclusion}
We have derived a new neural architecture to tackle the reconstruction of a dynamical process from partial and noisy observations. Both state trajectory and stochastic prior parametrization are learnt so that we also provide uncertainty quantification of the mean state. In this work, we use an SPDE-driven GP as flexible prior whose parameters are added as latent variables in an augmented state. A bi-level optimization scheme is used on both the inner variational cost derived from OI-based formulations and the outer training loss function of the neural architecture, which drives the optimization of the LSTM-based residual solver leading to the reconstruction of the state.

\clearpage

%\nocite{*}
\bibliographystyle{abbrvnat}
\bibliography{refimta}

%%%%%%%%%%%%%%%%%%%%%%%%%%%%%%%%%%%%%%%%%%%%%%%%%%%%%%%%%%%%
\section*{Checklist}

\iffalse
% %%% BEGIN INSTRUCTIONS %%%
The checklist follows the references. For each question, choose your answer from the three possible options: Yes, No, Not Applicable.  You are encouraged to include a justification to your answer, either by referencing the appropriate section of your paper or providing a brief inline description (1-2 sentences). 
Please do not modify the questions.  Note that the Checklist section does not count towards the page limit. Not including the checklist in the first submission won't result in desk rejection, although in such case we will ask you to upload it during the author response period and include it in camera ready (if accepted).
\textbf{In your paper, please delete this instructions block and only keep the Checklist section heading above along with the questions/answers below.}
% %%% END INSTRUCTIONS %%%
\fi

 \begin{enumerate}

 \item For all models and algorithms presented, check if you include:
 \begin{enumerate}
   \item A clear description of the mathematical setting, assumptions, algorithm, and/or model. [Yes/No/Not Applicable].  \textit{Yes, more specifically in Sections 2 and 3}
   \item An analysis of the properties and complexity (time, space, sample size) of any algorithm. [Yes/No/Not Applicable] \textit{Yes, when useful, e.g. complexity of OI with sparse precision matrices}
   \item (Optional) Anonymized source code, with specification of all dependencies, including external libraries. [Yes/No/Not Applicable] \textit{Yes, the GitHub repository will be provided in the final version of the paper}
 \end{enumerate}

 \item For any theoretical claim, check if you include:
 \begin{enumerate}
   \item Statements of the full set of assumptions of all theoretical results. [Yes/No/Not Applicable] \textit{Yes, see Section 2 on the SPDE formulation of the prior}
   \item Complete proofs of all theoretical results. [Yes/No/Not Applicable] \textit{Yes, see the Supplementary materials}
   \item Clear explanations of any assumptions. [Yes/No/Not Applicable] \textit{Yes, Sections 2 and 3}      
 \end{enumerate}

 \item For all figures and tables that present empirical results, check if you include:
 \begin{enumerate}
   \item The code, data, and instructions needed to reproduce the main experimental results (either in the supplemental material or as a URL). [Yes/No/Not Applicable] \textit{Yes, the GitHub repository will be provided in the final version of the paper. As outputs of the main code, a NetCDF file is produced, the same used to fill in the Table and produce the figures}
   \item All the training details (e.g., data splits, hyperparameters, how they were chosen). [Yes/No/Not Applicable] \textit{Yes, see Section 5: experimental settings}
         \item A clear definition of the specific measure or statistics and error bars (e.g., with respect to the random seed after running experiments multiple times). [Yes/No/Not Applicable] [Yes/No/Not Applicable] \textit{Yes, in Section 5 we provide the reference of the data challenge where the main metrics are precisely defined}
         \item A description of the computing infrastructure used. (e.g., type of GPUs, internal cluster, or cloud provider). [Yes/No/Not Applicable] \textit{Yes, see Section 5: experimental settings}
 \end{enumerate}

 \item If you are using existing assets (e.g., code, data, models) or curating/releasing new assets, check if you include:
 \begin{enumerate}
   \item Citations of the creator If your work uses existing assets. [Yes/No/Not Applicable] \textit{Yes, see Section 5 in which we provide the url of the data challenge related to our experiment}
   \item The license information of the assets, if applicable. [Yes/No/Not Applicable] \textit{Not applicable}
   \item New assets either in the supplemental material or as a URL, if applicable. [Yes/No/Not Applicable]  \textit{Yes, the GitHub repository will be provided in the final version of the paper.}
   \item Information about consent from data providers/curators. [Yes/No/Not Applicable] \textit{Not applicable}
   \item Discussion of sensible content if applicable, e.g., personally identifiable information or offensive content. [Yes/No/Not Applicable] \textit{Not applicable}
 \end{enumerate}

 \item If you used crowdsourcing or conducted research with human subjects, check if you include:
 \begin{enumerate}
   \item The full text of instructions given to participants and screenshots. [Yes/No/Not Applicable] \textit{Not applicable}
   \item Descriptions of potential participant risks, with links to Institutional Review Board (IRB) approvals if applicable. [Yes/No/Not Applicable] \textit{Not applicable}
   \item The estimated hourly wage paid to participants and the total amount spent on participant compensation. [Yes/No/Not Applicable] \textit{Not applicable}
 \end{enumerate}

 \end{enumerate}

\end{document}